# PND-Net: Physics based Non-local Dual-domain Network for Metal Artifact Reduction

Jinqiu Xia[1], Yiwen Zhou[1], Hailong Wang[1], Wenxin Deng[1], Jing Kang[1], Wangjiang Wu[1], Mengke Qi[1], Linghong Zhou[1], Jianhui Ma[2,*], Yuan Xu[1,*]

[1]School of Biomedical Engineering, Southern Medical University, Guangzhou 510515, China
[2]Department of Radiation Oncology, Nanfang Hospital, Southern Medical University, Guangzhou 510515, China

*Corresponding authors: jianhuima37@163.com and yuanxu@smu.edu.cn

**Abstract**

Metal artifacts caused by the presence of metallic implants tremendously degrade the reconstructed computed tomography (CT) image quality, affecting clinical diagnosis or reducing the accuracy of organ delineation and dose calculation in radiotherapy. Recently, deep learning methods in sinogram and image domains have been rapidly applied on metal artifact reduction (MAR) task. The supervised dual-domain methods perform well on synthesized data, while unsupervised methods with unpaired data are more generalized on clinical data. However, most existing methods intend to restore the corrupted sinogram within metal trace, which essentially remove beam hardening artifacts but ignore other components of metal artifacts, such as scatter, non-linear partial volume effect and noise. In this paper, we mathematically derive a physical property of metal artifacts which is verified via Monte Carlo (MC) simulation and propose a novel physics based non-local dual-domain network (PND-Net) for MAR in CT imaging. Specifically, we design a novel non-local sinogram decomposition network (NSD-Net) to acquire the weighted artifact component, and an image restoration network (IR-Net) is proposed to reduce the residual and secondary artifacts in the image domain. To facilitate the generalization and robustness of our method on clinical CT images, we employ a trainable fusion network (F-Net) in the artifact synthesis path to achieve unpaired learning. Furthermore, we design an internal consistency loss to ensure the data fidelity of anatomical structures in the image domain, and introduce the linear interpolation sinogram as prior knowledge to guide sinogram decomposition. NSD-Net, IR-Net, and F-Net are jointly trained so that they can benefit from each other. Extensive experiments on simulation and clinical data demonstrate that our method outperforms the state-of-the-art MAR methods.

## I. INTRODUCTION

Computed tomography (CT) can noninvasively reveal the human anatomy and has been widely used for medical diagnosis and assessment. In radiation therapy, CT image is generally employed for organ delineation, including the gross tumor volume (GTV) and surrounding organs at risk (OARs). Additionally, treatment planning requires accurate electron density related to the Hounsfield Units (HU) of CT images to calculate dose distribution [1]. Advances in surgical treatments lead to the metallic implants frequently appearing in patients, such as dental fillings, prosthesis and cardiac pacemaker. However, when polychromatic X-ray beams penetrate high-density implants, beam hardening, scatter, noise and non-linear partial volume effect are prominently generated, which are collectively referred to as metal artifacts. As a result, the corrupted X-ray projections (sinogram) will be utilized to solve the ill-posed inverse problem, subsequently causing severe bright and dark streaking artifacts in reconstructed CT images. These metal artifacts not only affect clinical diagnosis, but also increase errors of organ delineation and dose calculation in radiation therapy. Thus, metal artifact reduction (MAR) has become a critical task in CT imaging.

In the past decades, many MAR methods have successively been proposed [2], which can be roughly categorized into three groups: iterative reconstruction [3, 4], sinogram correction [5-8], and image correction [9]. The MAR methods corrected in image domain, also named post-processing methods, cannot reduce dark and bright bands around metallic implants since metal artifacts are non-local and structured in the reconstructed images. The basic principle of iterative reconstruction methods is to estimate the error between measured sinogram and calculated sinogram, and then update the current reconstruction in terms of the sinogram error. This process is repeated a number of times until an objective function is optimized, resulting in expensive computational cost which restricts the widespread application of MAR iterative approaches in clinic. In addition, the sinogram correction methods, including interpolation-based methods [5, 10], prior image-based methods [11-13] and physics-based methods [14, 15] are widely used in practice. The linear interpolation metal artifact reduction (LIMAR) approach [5] restored the metal affected regions by the linear interpolation of adjacent uncorrupted data along the detector channel direction of sinogram. However, LIMAR usually introduces new artifacts in the anatomical structures surrounding metallic implants. To overcome the inconsistency between interpolated values and measured values in the sinogram, prior image-based methods such as normalized MAR (NMAR) [13], tissue-class method [12] and prior image fusion method [11] were developed. By re-projecting prior images, these methods can better estimate the missing projections corrupted by the metal trace. Besides, some methods [16-18], which model the latent physical effects of metal artifacts, were proposed to correct metal artifacts by dealing with beam hardening and photon starvation.

With the development of deep learning, the convolutional neural network (CNN) has been applied for MAR based on image processing and sinogram completion. The supervised methods [19-26] performed well on synthesized paired CT images. Park et al. [17] utilized a U-Net [27] to restore inconsistent sinogram by reducing beam hardening artifacts. Yu et al. [24] first established a metal-free and metal-inserted dataset and proposed a prior image-based method. The network took uncorrected and pre-corrected images as input and then generated a CNN prior image for final sinogram correction. However, above methods only deal with metal artifacts in image domain or sinogram domain, limiting the network performance. Therefore, end-to-end dual-domain networks [21-23] were successively proposed and outperformed single domain based methods [17, 24, 28]. Lin et al. developed the first dual-domain refinement network DuDoNet [19], which used a sinogram network to enhance the output of image network. Recently, Li et al. [26] proposed a Fourier dual-domain network for MAR. A Fourier sinogram restoration network and a Fourier refinement network in the image domain were jointly utilized to refine the reconstructed image in a local-to-global manner. Although the supervised methods show reasonable results on MAR, they heavily rely on paired metal-free and metal-corrupted CT images for network training, which are hard to be acquired in clinic. Furthermore, since the measured metal artifacts are much more complicated than the synthesized artifacts, the supervised methods cannot generalize well on real clinical data. Consequently, unsupervised methods trained with unpaired data had been proposed to solve MAR problem [29-32]. For example, Liao et al. proposed AND [29] composed of encoder and decoder architecture, which adopted unsupervised learning for image-to-image translation. However, without the correction in sinogram domain, ADN still remains shading artifacts nearby metallic implants. So the U-DuDoNet [30] employed dual-domain correction and unpaired data to achieve a good performance on clinical data compared with other unsupervised methods. It jointly constructed two U-Nets in sinogram and image domains, and introduced a self-learned sinogram inpainting network to enhance sinogram completion. Moreover, Yu et al. [33] trained a dual-domain network via cross-domain learning to conduct self-supervised MAR, which can also lessen dependence on paired data.

Aforementioned dual-domain methods [19, 20, 26, 30, 33] generally utilizing a sinogram completion network to restore missing projections within the metal trace, also known as inpainting methods [30, 33]. However, metal artifacts are mainly caused by beam hardening, scatter, non-linear partial volume effect and noise, where beam hardening artifacts are confined to the metal trace region but other artifact components are non-local in the sinogram domain. Therefore, previous methods only remove beam hardening artifacts, which would limit the capability of suppressing metal artifacts.

In this work, we mathematically derive the physical property of metal artifacts and propose a novel physics based non-local dual-domain network (PND-Net) for MAR task. To validate this physical property of metal artifacts, Monte Carlo (MC) simulation is employed to simulate the transport process of polychromatic X-ray beams penetrating a phantom with metallic implants. Different from previous methods that restore the corrupted data within metal trace in the sinogram domain, we propose to decompose the artifact components including beam hardening, scatter, non-linear partial volume effect and noise from the whole sinogram. Then, a non-local sinogram decomposition network (NSD-Net) is designed to reduce the sinogram artifact component. In the image domain, we use an image restoration network (IR-Net) to refine the residual and secondary artifacts in the reconstructed CT image while preserving the normal tissue details. Unpaired learning is implemented in a cyclic artifact separation and synthesis process. Unlike U-DuDoNet [30] which is based on additive property, we design a fusion network (F-Net) to synthesize the metal-corrupted image, which can contribute to the better final MAR result. Specifically, we introduce the LI sinogram as prior knowledge for NSD-Net to decompose sinogram artifact components, and propose an internal consistency loss to preserve structural textures in the reconstructed CT image. Extensive experiments demonstrate that our proposed PND-Net outperforms the state-of-the-art unsupervised methods on simulation data, and achieve a visually superior image quality compared with other supervised approaches on clinical data. Our main contributions are summarized as follows:

1. Based on the physical property that beam hardening dominates metal artifacts but other artifact components are not only restricted to the metal trace region in the sinogram domain, we design a novel non-local sinogram decomposition network to enhance sinogram refinement. The proposed network utilizes the non-local weight block to yield a weight map which can declare the priority of artifact components in the sinogram domain.

2. We adopt dual path learning to facilitate the generalization of our framework and build a trainable fusion network in synthesis path for superior metal artifact reduction.

3. We propose an internal consistency (IC) loss to ensure the data fidelity of anatomical structures in the image domain. The IC loss can enforce IR-Net for better preservation of normal tissue details in the final MAR image of our method.

4. We introduce the linear interpolation sinogram as prior loss to guide sinogram decomposition.

The remainder of this paper is organized as follows. We introduce our method in Section II. The dataset and evaluation metrics are presented in Section III. The experimental results are compared qualitatively and quantitatively in Section IV. The strengths and limitations are discussed and a conclusion is drawn in Section V.

## II. METHODOLOGY

### A. Physical property of metal artifacts

The CT image intensity represents linear attenuation coefficient of the anatomical structure. Let $X = \mu(x, E)$ denote a linear attenuation coefficient slice of metal-free human body at position $x$ with energy $E$, $\mathcal{P}(\cdot)$ indicates forward projection operator. According to Lambert-Beer law, the sinogram $\mathcal{P}(X)$ of polychromatic X-ray beams can be expressed as

$$\mathcal{P}(X) = -\ln\{\int \eta(E) \exp(-\int_L \mu(x, E) dl) dE\} \quad (1)$$

where $\eta(E)$ denotes the weight of normalized spectrum energy, and $L$ is the path from source to detector. When metals are implanted in the imaging region, the metal-inserted image $X^m$ can be formulated as $X^m = X \odot (1 - M) + \mu_{metal}(E) \odot M$, where $M \in \{0, 1\}$ denotes a binary metal mask, $\odot$ and $\mu_{metal}(E)$ represent element-wise multiplication and linear attenuation coefficient of metallic implants, respectively. The metal-affected sinogram $\mathcal{P}(X^m)$ can be then defined as

$$\begin{aligned}\mathcal{P}(X^m) &= -\ln\{\int \eta(E) \exp(-\int_L [\mu(x,E) \odot (1-M) + \mu_{metal}(E) \odot M] dl) dE\} \\ &= \mathcal{P}(X) - \ln\{\int \eta(E) \exp(-\mu_{metal}(E) \int_L M \ dl) dE\}\end{aligned} \quad (2)$$

Ideally, an accurate CT image reconstructed from sinogram $\mathcal{P}(X)$ or $\mathcal{P}(X^m)$ needs to satisfy that the normalized spectrum energy $\eta(E)$ is constant when photons propagate in the human body [34]. However, $\eta(E)$ varies with different energy level $E$. Since lower energy photons are more likely to be absorbed by metal materials, high energy photons gradually dominate the normalized spectrum energy. As a result, beam hardening process occurs when polychromatic X-rays penetrate metallic implants. Therefore, the beam hardening artifacts $a_{BH}$ can be formulated as:

$$a_{BH} = \mathcal{P}(X^m) - \mathcal{P}(X) = -\ln\{\int \eta(E) \exp(-\mu_{metal}(E)\mathcal{P}(M)) dE\} \quad (3)$$

where metal projection $\mathcal{P}(M) = \int_L M dl$. Obviously, beam hardening artifacts only locate within the nonzero region which is called metal trace $M_t$, and $M_t = \delta(\mathcal{P}(M))$ where $\delta(\cdot)$ is a binary indicator function [30].

At present, most state-of-the-art methods [20, 23, 26, 30, 35] intend to restore the corrupted sinogram within metal trace $M_t$, which essentially remove beam hardening artifacts but circumvent other components of metal artifacts. Nevertheless, we validate that other metal artifacts containing scatter, non-linear partial volume effect and noise are not only restricted to the metal trace region via MC simulation [36] and real projections, and these artifacts are inevitable in MAR.

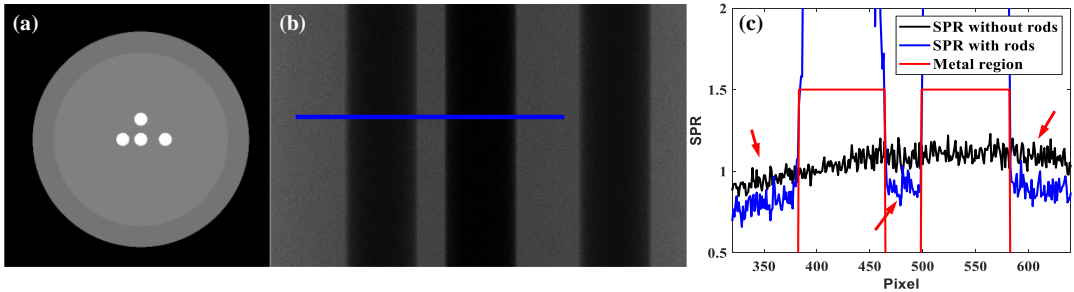

**Fig. 1.** The water phantom inserted four titanium rods via MC simulation is shown in (a). (b) is the primary signal at projection angle $0°$, and (c) shows the scatter-to-primary ratio profiles of the phantom with and without rods along the blue line in (b).

Fig. 1(a) shows the water phantom inserted with four metal rods whose resolution is $512 \times$

$512 \times 130$ with a voxel size of $0.49 \times 0.49 \times 3\ mm^3$. The primary projection is simulated with $10 \times 10^{10}$ source photons for beam energy of 1 to 100 keV, and the resolution of projection image is $1024 \times 768$ with a pixel size of $0.6 \times 0.6\ mm^2$. The distances from the X-ray source to the isocenter (SID) and detector (SDD) are separately 100 cm and 150 cm. We employ an in-house MC simulation tool [36] to generate primary and scatter signals of the water phantom with and without metal rods, respectively. The scatter-to-primary ratio (SPR) profiles are drawn along the blue line to show the impact of metal rods. As shown in Fig. 1(c), the SPR around metal rods is apparently decreased compared to the water phantom without metals, suggesting that metals can also cause changes in scattering besides beam hardening.

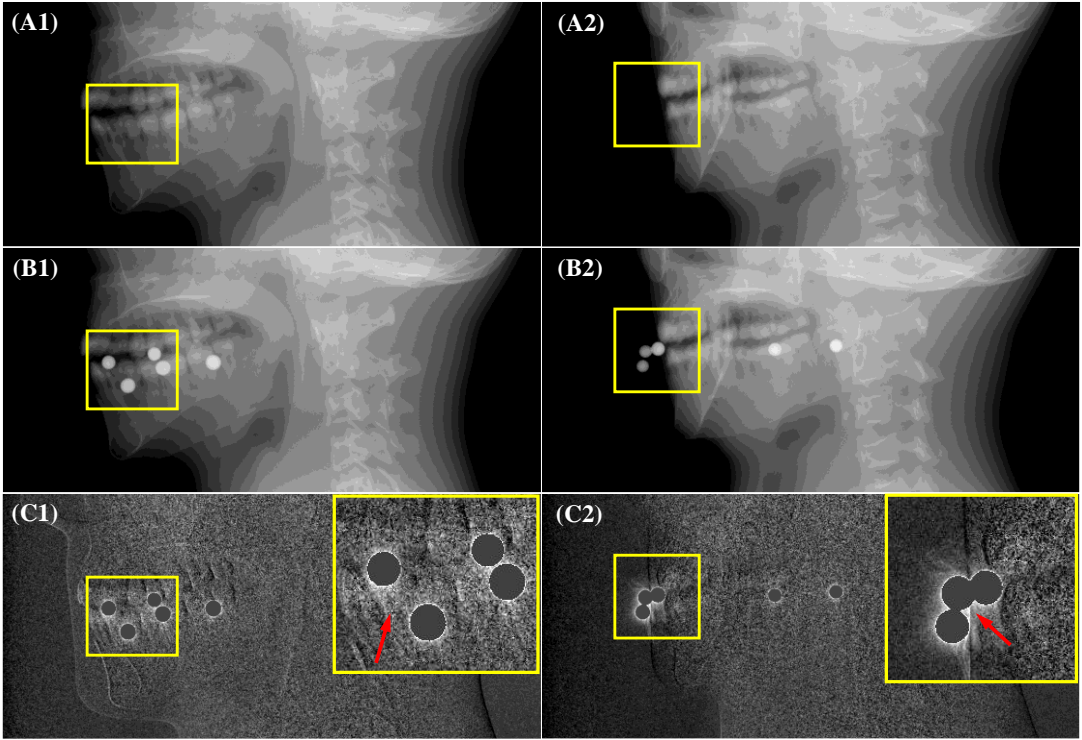

**Fig. 2.** The influence of metals in realistic Cone beam CT projections of a head phantom. The projections at different angles with and without metal implants are displayed in columns. (C1) and (C2) are the differences between (A1-A2) and (B1-B2). The values of metal region are set to zero for better observation.

Fig. 2 shows the influence of metal implants in the real projection. Two different projections of the head phantom with and without metal implants are acquired, respectively. The differences between (A1-A2) and (B1-B2) are mainly caused by scatter, non-linear partial volume effect and noise [1, 2]. As shown in Fig. 2(C1) and (C2), the intensity around metals, which indicated by red arrows, changes due to the presence of the metal, implying that metal artifacts are non-local in sinogram domain.

Since MC simulation and real data analysis demonstrate that the effect of metal artifacts peripheral to the metal trace region in sinogram domain is not negligible, we formulate MAR as a non-local sinogram decomposition problem, which further takes scatter $a_{scatter}$, non-linear partial volume effect $a_{NPVE}$ and noise $a_{noise}$ into account:

$$S_a = a_{BH} + a_{scatter} + a_{NPVE} + a_{noise} \qquad (4)$$

$S_a$ indicates non-local metal artifacts in the sinogram domain. Then the corrected sinogram $S_{mc}$ can be obtained by subtracting $S_a$ from metal-affected sinogram $S_m$:

$$S_{mc} = S_m - S_a \qquad (5)$$

Additionally, the filtered back projection (FBP) $\mathcal{P}^{-1}(\cdot)$ is a linear operator. Thus, the MAR processing (5) can be described as

$$I_{mc} = \mathcal{P}^{-1}(S_m - S_a) = I_m - I_a \qquad (6)$$

where $I_a$ is metal artifacts in the reconstructed CT image, $I_{mc}$ and $I_m$ denote MAR CT image and metal-corrupted image, respectively.

When the non-local metal artifacts containing $a_{BH}$, $a_{scatter}$, $a_{NPVE}$ and $a_{noise}$ are modeled, the MAR processing can be expressed as (5) or (6), which is distinguished from previous methods [22, 23, 26, 30, 35]. Based on this property, we design a metal artifact decomposition and synthesis framework and construct our networks as follows.

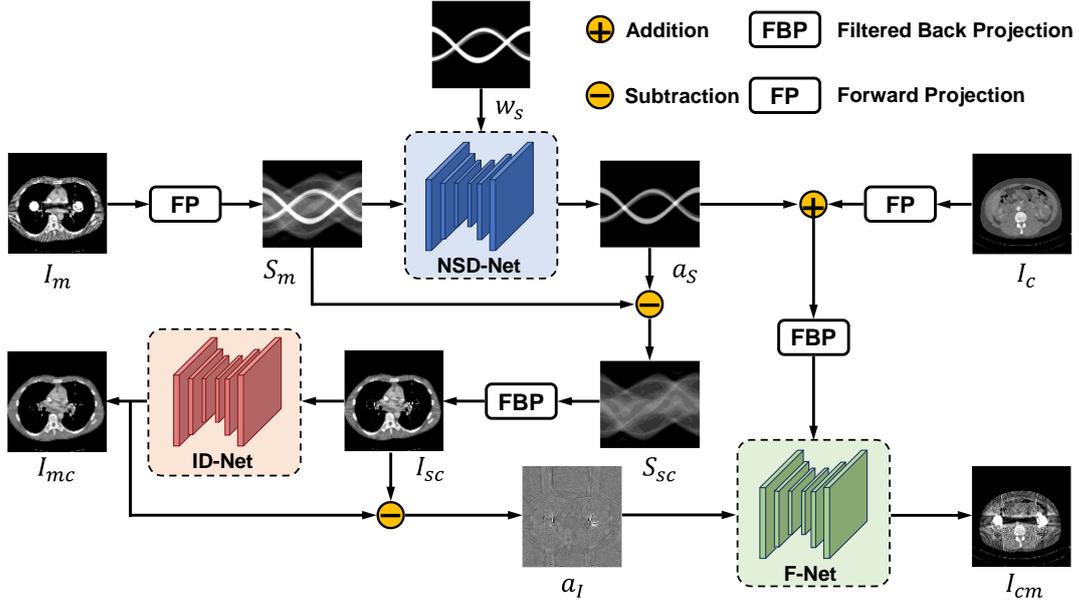

**Fig. 3.** The proposed physics based non-local dual-domain network (PND-Net) for metal artifact reduction. Given the corrupted sinogram $S_m$ and weight matrix $w_s$, NSD-Net generates the weighted artifact component $a_S$ with LI sinogram as prior loss. The corrected sinogram $S_{sc}$ can be obtained by subtracting $a_S$ from $S_m$. Then IR-Net is employed to remove the residual artifacts $a_I$ to acquire the final MAR image $I_{mc}$ with the guidance of IC loss. In artifact synthesis path, F-Net takes $a_I$ and the sinogram corrupted image as inputs to produce a reasonable synthesized metal-corrupted image $I_{cm}$. NSD-Net, IR-Net, and F-Net are jointly trained to benefit from each other. In testing phase, given a metal-corrupted CT image $I_m$, PND-Net can output the MAR image $I_{mc}$.

### B. Framework Overview

Fig. 3 shows the proposed PND-Net framework. In terms of the physical property of metal artifacts, we first design a non-local sinogram decomposition network (NSD-Net) to reduce beam hardening, scatter and NPVE that induce dark bands and bright borders in the reconstructed CT images. Different from inpainting methods, we feed a metal-corrupted sinogram $S_m$ and an artifact weight matrix $w_s$ into our NSD-Net to generate the sinogram artifact component $a_S$ and acquire a corrected sinogram $S_{sc}$. To further alleviate residual artifacts which cannot be completely suppressed in the sinogram domain, we simultaneously develop an image restoration network (IR-Net) to acquire the final MAR image $I_{mc}$ by refining the sinogram corrected image $I_{sc}$, which is reconstructed from $S_{sc}$ with the conventional FBP algorithm. Furthermore, considering that paired metal-free and metal-corrupted CT images are hard to be acquired in clinic, we employ unpaired learning with a synthesis path to enhance the generalization and robustness of our framework. Since the residual artifact $a_I$ is closely related to the anatomical structures and contains texture information, it is not plausible to synthesize the metal-corrupted image $I_{cm}$ by directly adding $a_I$ to an arbitrary clean CT image $I_c$. Therefore, we utilize a fusion network (F-Net) to produce a reasonable synthesized image. F-Net takes $a_I$ and the sinogram corrupted image synthesized by the sinogram artifact component $a_S$ and forward projection of the clean image $I_c$ as inputs, where the residual artifact $a_I$ can be obtained by subtracting the final MAR image $I_{mc}$ from $I_{sc}$. NSD-Net, IR-Net, and F-Net are jointly trained so that they can be mutually beneficial to achieve non-local sinogram decomposition, image restoration, and artifact synthesis processes.

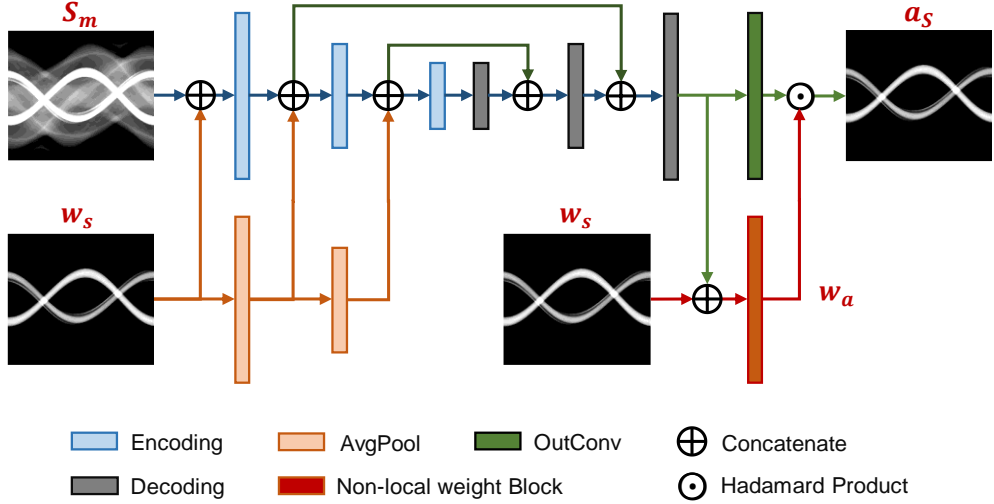

**Fig. 4.** Non-local sinogram decomposition network architecture.

## B.1. Non-local Sinogram Decomposition

Since beam hardening dominates metal artifacts but other artifact components are not only restricted to the metal trace region in the sinogram domain, we propose a non-local sinogram decomposition network NSD-Net with taking a metal-corrupted sinogram $S_m$ and a weight matrix $w_s$ as inputs to generate the weighted artifact component $a_S$. The weight matrix declares the priority of artifact components in the sinogram domain, which generated by $w_s = \mathcal{P}(M) \odot M_t + \varepsilon \odot (1 - M_t)$. Specifically, $\mathcal{P}(M)$ is mainly affected by beam hardening, while non-metal trace region $1 - M_t$ weighted with a nonzero minor value $\varepsilon = 0.001$ represents scatter and non-linear partial volume effect. As shown in Fig. 4, we choose mask pyramid U-Net [37] with three encoding and decoding blocks as backbone, and $S_m$ concatenated with $w_s$ is encoded at first. $w_s$ is then downsampled by average pooling and concatenate $S_m$ before each encoding layer. Inspired by the first-order model of beam hardening compensation [18], we design a non-local weight block composed of three convolutional layers with kernel sizes of $1 \times 1$, $3 \times 3$ and $1 \times 1$ to predict the weighted artifact map $w_a$. The first two convolutional layers are followed with an instance normalization and a leaky rectified linear unit (LeakyReLU) function, and a Tanh function is utilized as the activation function of the last convolutional layer. The feature maps generated from three decoding layers are concatenated with $w_s$ to be fed into the non-local weight block. Then, we use Hadamard product to multiply $w_a$ and the feature maps of OutConv, which is a $1 \times 1$ convolutional layer with one output channel, to produce the final weighted artifact component $a_S$. Finally, the non-local corrected sinogram $S_{sc}$ can be acquired by subtracting $a_S$ from original

sinogram $S_m$:
$$S_{sc} = S_m - f_S(S_m, w_s) \tag{7}$$
where $f_S$ represents NSD-Net and $a_S = f_S(S_m, w_s)$.

**B.2. Image Artifact Reduction**

Sinogram domain processing cannot completely eliminate metal artifacts, and FBP operation would introduce secondary artifacts due to the lack of geometry consistency in sinogram correction. Therefore, we employ IR-Net, which is a residual U-Net [38] of depth 4, to address the residual artifacts mainly arising from noise and secondary artifacts in the image domain. The sinogram corrected image $I_{sc}$ reconstructed from $S_{sc}$ is fed into IR-Net to output the final MAR image $I_{mc}$. So the image artifact reduction can be expressed as:
$$I_{mc} = f_I(\mathcal{P}^{-1}(S_{sc})) \tag{8}$$
where $f_I$ denotes IR-Net, and the image artifact $a_I$ can be acquired via subtracting $I_{mc}$ from $I_{sc}$, that is $a_I = I_{sc} - I_{mc}$.

**B.3. Synthesized Artifact-contaminated Image**

A data-driven deep neural network heavily requires paired data which is a demanding task in clinic. To improve the generalization of our framework and effectively exploit clinical data, we add an artifact synthesis path to achieve unpaired learning. In addition to generating the MAR image $I_{mc}$, we also synthesize a new metal artifact-contaminated CT image $I_{cm}$. Different from U-DuDoNet that directly adds artifact components $a_S$ and $a_I$ to a clean CT image $I_c$, we develop a F-Net for the artifact synthesis process. The addition of the arbitrary CT image and the image artifact which contains anatomical textures may lead to mode collapse, while the joint training of F-Net and IR-Net can stabilize unpaired learning. Therefore, we formulate the artifact synthesis as
$$I_{cm} = f_F(\mathcal{P}^{-1}(\mathcal{P}(I_c) + a_S), a_I) \tag{9}$$
F-Net denoted as $f_F$ is a mask pyramid U-Net [37], where the depth is 4 and the filters is 32 at first. And the average pooling is utilized to down sample $a_I$ which then concatenates the feature maps of the decoder.

**C. Overall Objective Function**

In training phase, our framework takes two unpaired CT images, i.e., a metal-affected image $I_m$ and a clean CT image $I_c$ as inputs, the overall objective function contains four losses: (a) adversarial losses, (b) cycle loss, (c) LI prior loss and (d) IC loss. Adversarial loss and cycle loss are utilized for unpaired learning, which imitate CycleGAN [39] training style. In addition, we design LI prior loss and IC loss to improve the network performance. We introduce the LI corrected result as prior sinogram to guide the sinogram completion. To ensure the data fidelity in the image domain, an IC loss is proposed for stabilizing the network training on real clinical data. In testing phase, given a metal-corrupted CT image $I_m$, our PND-Net can output the MAR image $I_{mc}$.

**Adversarial Loss.** Learning to generate the corrected image $I_{mc}$ and synthesize the artifact-contaminated image $I_{cm}$ is a crucial task for our method. However, since there are no paired images, we apply two discriminators to achieve adversarial learning [40]. Specifically, discriminator $D^c$ aims to distinguish between the real clean image $I_c$ and the MAR image $I_{mc}$, while discriminator $D^m$ tries to differentiate the synthesized image $I_{cm}$ from the real artifact-affected image $I_m$. The adversarial loss can be expressed as

$$\mathcal{L}_{adv} = \mathbb{E}[\log D^c(I_c)] + \mathbb{E}[1 - \log D^c(I_{mc})] + \mathbb{E}[\log D^m(I_m)] + \mathbb{E}[1 - \log D^m(I_{cm})] \quad (10)$$

**Cycle Loss.** During unpaired training, decomposition and synthesis processes in our framework is cyclic. When the synthesized corrupted image $I_{cm}$ and corrected image $I_{mc}$ are fed back to PND-Net, cyclic reconstructed images $I'_c$ and $I'_m$ can be acquired. Note that $I'_m$ and $I'_c$ should be consistent with the original images $I_c$ and $I_m$, respectively. We use $L_1$ loss to minimize the distance between the original and cyclic reconstructed images, the cyclic reconstruction loss can be written as

$$\mathcal{L}^{rec}_{cycle} = \|I_m - I'_m\|_1 + \|I_c - I'_c\|_1 \quad (11)$$

Similarly, the cyclic artifact components $a'_S$ and $a'_I$ extracted from the synthesized image $I_{cm}$ should also be identical with $a_S$ and $a_I$, the cyclic artifact loss is then formulated as

$$\mathcal{L}^{art}_{cycle} = \|a_S - a'_S\|_1 + \|a_I - a'_I\|_1 \quad (12)$$

Therefore, the cycle loss is the sum of $\mathcal{L}^{rec}_{cycle}$ and $\mathcal{L}^{art}_{cycle}$.

**LI Prior Loss.** To guide the artifact decomposition in the sinogram, we introduce the simple but robust method LIMAR to obtain the LI corrected result $S_{LI}$ as a prior sinogram. However, directly using LI corrected sinogram will introduce secondary artifacts. To mitigate the inconsistency between interpolated values and neighboring unaffected values in the sinogram, we use a Gaussian blur operation $\mathcal{H}_{\sigma=3}$ with $\sigma = 3$ and $L_1$ loss to minimize the distance between the blurred LIMAR sinogram and blurred $S_{sc}$:

$$\mathcal{L}_{LI} = \|\mathcal{H}_{\sigma=2}(S_{sc}) - \mathcal{H}_{\sigma=2}(S_{LI})\|_1 \quad (13)$$

**Internal Consistency Loss.** In our method, IR-Net is employed to solve the image restoration problem. IR-Net aims to reduce the image artifacts without the loss of normal tissues. That is, the output should be equal to the input, when a clean CT image $I_c$ is fed into IR-Net. Therefore, we employ the $L_1$ loss to ensure the data fidelity in the image domain:

$$\mathcal{L}_{IC} = \|f_I(I_c) - I_c\|_1 \quad (14)$$

In summary, the total objective function is the weighted sum of the above losses:

$$\mathcal{L}_{total} = \mathcal{L}_{adv} + \alpha_1\left(\mathcal{L}^{art}_{cycle} + \mathcal{L}^{rec}_{cycle}\right) + \alpha_2 \mathcal{L}_{LI} + \alpha_3 \mathcal{L}_{IC} \quad (15)$$

where we empirically set $\alpha_1 = 20.0$, $\alpha_2 = 8$ and $\alpha_3 = 5$ in simulation data. For clinical data, we set $\alpha_1 = 30$, $\alpha_2 = 20$ and $\alpha_3 = 20$.

## III. EXPERIMENTS

### A. Simulation and Clinical Dataset

We evaluated our method on both simulation and real clinical data. For simulation data, we synthesized metal-affected CT images following the method [23, 27, 29-31]. Specifically, we randomly selected 1000 clean CT images and 90 metal masks from a public dataset named DeepLesion [41] for training, and other 250 CT images from 15 patients with the remaining 10 metal masks for testing. The CT images were resized to $256 \times 256$ and the sinogram was generated with 320 projection views and 641 detector bins in a row under the protocol of fan-beam scanning. The distances from the X-ray source to the isocenter (SID) and detector (SDD) are separately 107.5 cm and 150 cm. For real clinical data, we retrospectively collected 400 metal-corrupted images and 420 clean CT images from 32 patients. 400 metal-corrupted images were randomly split into 350 images for training and 50 images for testing. The metal implants were segmented with a threshold of 2500 HU. We cropped and resized the clinical images to the same size of simulation data. The above data pre-processing was implemented with ASTRA Toolbox [42].

### B. Implementation Details

Our model was implemented in PyTorch and trained on a desktop workstation equipped with a NVIDIA TITAN Xp graphics processing unit (GPU) with 12 gigabytes of memory. The FP and FBP operations were implemented with ODL library (https://odlgroup.github.io/odl). We trained our model 300 epochs using the Adam optimizer [43] with $(\beta 1, \beta 2) = (0.5, 0.999)$. The initial learning rate was $2 \times 10^{-4}$ and the mini-batch size was 2. For two discriminators, we use the other two Adam optimizers with the same learning rate of $1 \times 10^{-4}$. The code for reproducing this study is available at https://github.com/Ballbo5354/PND-Net.

### C. Evaluation Metrics and Baselines

In this study, peak signal-to-noise ratio (PSNR), root mean square error (RMSE), and structural similarity index (SSIM) are employed to quantitatively evaluate the MAR image. We compare our method with different state-of-the-art (SOTA) MAR methods. Specifically, LIMAR [5] and NMAR [13] are traditional method. The supervised methods including U-Net [27] and DuDoNet++ [23] are trained with simulation data and tested on both simulation and clinical data, where U-Net takes the LI corrected CT image and uncorrected image as inputs. AND [29], CycleGAN [39] and U-DuDoNet [30] are unsupervised methods which are trained and tested on simulation data or clinical data. Since U-DuDoNet adopts a self-learned prior net (P-Net) trained with artificial metal-corrupted sinograms to enhance the sinogram completion, the pre-trained P-Net is closely related to the synthesized training pairs, we use LI corrected sinogram to replace the predictive uncertainty of P-Net. Therefore, U-DuDoNet is called U-DuDoNet-LI here.

## IV. RESULTS

### A. Experimental Results on Simulation Data

**Visual Comparison.** Fig. 5 shows the visual comparisons with different SOTA methods on DeepLesion simulation data, and the zoomed subfigure reveals image details around metallic implants. As shown in Fig. 5(B1-B3), three CT images of chest, abdomen, and pelvis suffer from severe dark and bright streaks, especially normal tissues nearby metallic implants. For traditional methods, it is observed that LIMAR and NMAR can remove obvious streaking artifacts (see Fig. 5(C1-C3 and D1-D3)). However, the discontinuity between restored data within metal-corrupted region and unaffected values in the sinogram introduces secondary artifacts and loses the details of the anatomical structure close to the implant. As for supervised methods, U-Net (Fig. 5(E1-E3)) can obtain better visual results than traditional algorithms, but still contains newly introduced artifacts (Fig. 5(E1)) and blurs the spine around the metal (Fig. 5(E2)). It is because it directly takes the blended information of uncorrected and LIMAR images as input. And the DuDoNet++ (Fig. 5(F1-F3)) performs much better than U-Net, particularly in abdomen case.

As shown in Fig. 5(G2-G3 and H2-H3), image domain-based methods (CycleGAN and AND) are not able to reduce dark bands completely in the region between two metallic implants, but can effectively alleviate bright borders surrounding the implants. The zoomed images of Fig. 5(I1-I3 and J1-J3) suggest that both U-DuDoNet-LI and our method PND-Net preserve more anatomical details than CycleGAN and AND, which can be attributed to the dual-domain correction. Furthermore, our method retains distinct structural edges and contains least streaking artifacts among all unsupervised MAR methods.

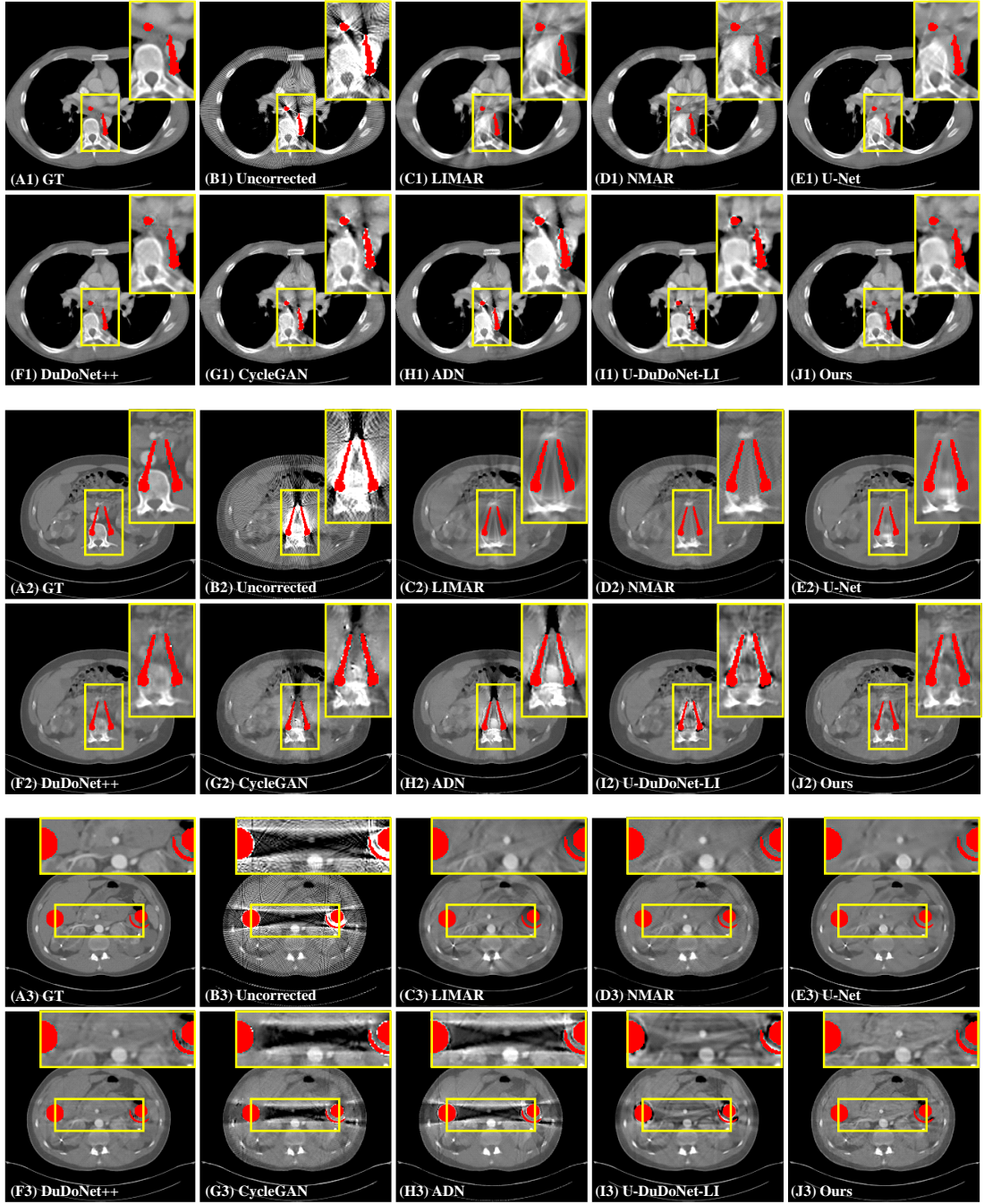

**Fig. 5.** Visual comparisons with different SOTA methods on simulation data. The groundtruth (A1-A3) and metal-corrupted images (B1-B3) of three cases are showed, respectively. We compare our method (J1-J3) with LIMAR (C1-C3), NMAR(D1-D3), U-Net (E1-E3), DuDoNet++ (F1-F3), CycleGAN (G1-G3), AND (H1-H3), and U-DuDoNet-LI (I1-I3). The display window is [-320 600] HU and metal implants are colored in red for better visualization.

**Quantitative Comparison.** From Table 1, we can observe that the quantitative improvement from U-Net to DuDoNet++ and from AND to U-DuDoNet-LI or our method, suggesting that dual-domain correction model outperforms image domain-based model. In traditional methods, LIMAR outperforms NMAR, since the prior image including pseudo bones or tissues will mislead to an incorrect segmentation during clustering. Among all unsupervised methods, our method achieves the best performance, where RMSE is decreased by 20.71, 15.14, and 9.05 compared with CycleGAN, AND, and U-DuDoNet-LI. Quantitatively compared with supervised approaches, our method is superior to U-Net in terms of RMSE and PSNR. Benefited from training with paired data in an end-to-end manner, DuDoNet++ attains the highest SSIM and PSNR, and lowest RMSE on simulation data.

Table 1. Quantitative comparison of different state-of-the-art methods on simulation data.

|  | Method | SSIM | RMSE (HU) | PSNR (dB) |
|---|---|---|---|---|
| Uncorrected | – | 0.321±0.049 | 112.23±26.54 | 20.80±2.26 |
| Traditional | LIMAR [5] | 0.884±0.0253 | 31.87±8.31 | 31.90±2.45 |
|  | NMAR [13] | 0.811±0.031 | 39.93±5.73 | 29.72±1.71 |
| Supervised | U-Net [27] | 0.957±0.012 | 28.60±4.37 | 32.49±1.67 |
|  | DuDoNet++ [23] | **0.978±0.005** | **15.10±3.17** | **38.18±2.17** |
| Unsupervised | CycleGAN [39] | 0.902±0.024 | 44.84±12.66 | 28.94±2.64 |
|  | ADN [29] | 0.908±0.018 | 39.27±14.86 | 30.42±3.31 |
|  | U-DuDoNet-LI [30] | 0.944±0.015 | 33.18±7.73 | 31.46±2.39 |
|  | PND-Net (Ours) | **0.949±0.010** | **24.13±6.23** | **34.01±2.06** |

**B. Experimental Results on Real Clinical Data**

Fig. 6 presents the visual comparison of different methods on three clinical cases with real metal artifacts. It is observed that conventional LIMAR and NMAR methods introduce severe bright streaking artifacts and distort the bone structures, as indicated in the zoomed subfigures. Instead, U-Net and DuDoNet++ can reduce more artifacts but introduce dark borders surrounding metallic implants and smooth out the whole CT image [19, 21]. Furthermore, DuDoNet++ synthesizes unreliable bone structures along the axis of greatest attenuation. As shown in Fig. 6(F1-F3 and G1-G3), soft tissues suffer from obvious dark bands since AND and DuDoNet-LI only recover the corrupted sinogram within the metal trace region. Compared with above SOTA methods, our method effectively restores the fine details of the anatomical structures and reduces most artifacts.

To further verify the performance of preserving structural details and reducing metal artifacts, we select two regions of interest (ROIs) in the clinical CT image to visually compare our method with different SOTA methods. As shown in Fig. 7(A1-H1), the bone structures are lost in the LIMAR and NMAR results, while U-Net and DuDoNet++ blur soft tissues and bones in the uncorrupted ROI. All unsupervised methods retain the details of the original structures but our method contains least streaks compared with AND and DuDoNet-LI, as indicated in Fig. 7(A2-H2).

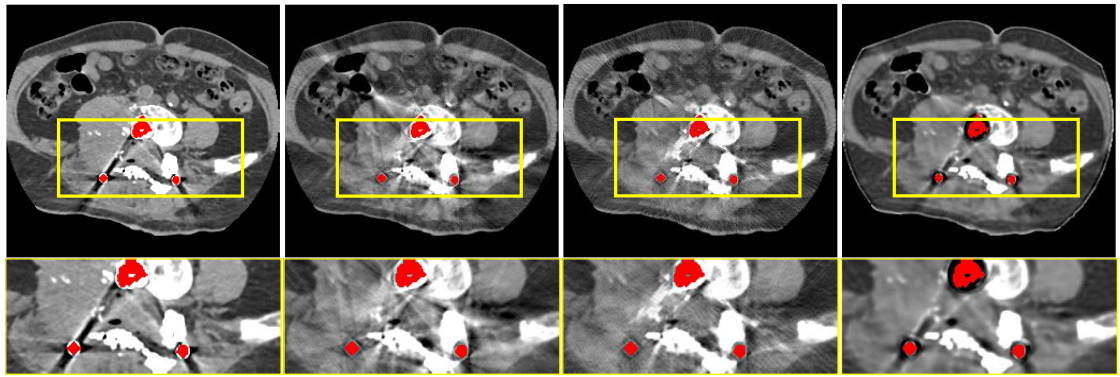
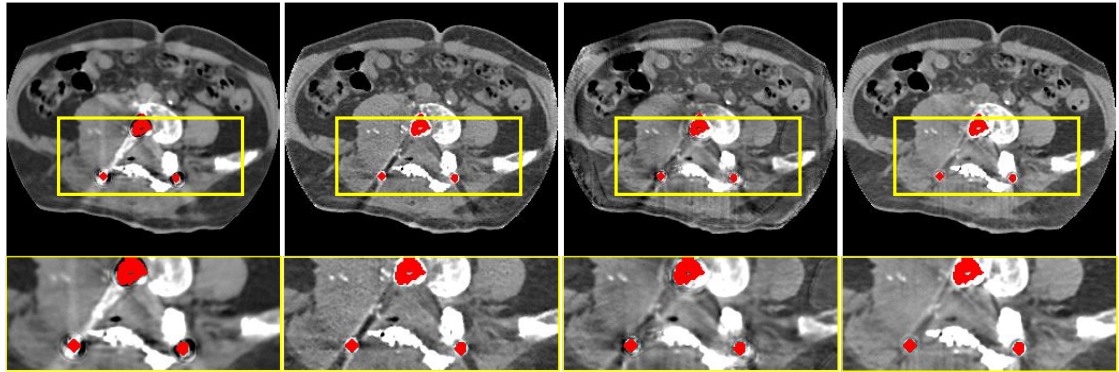
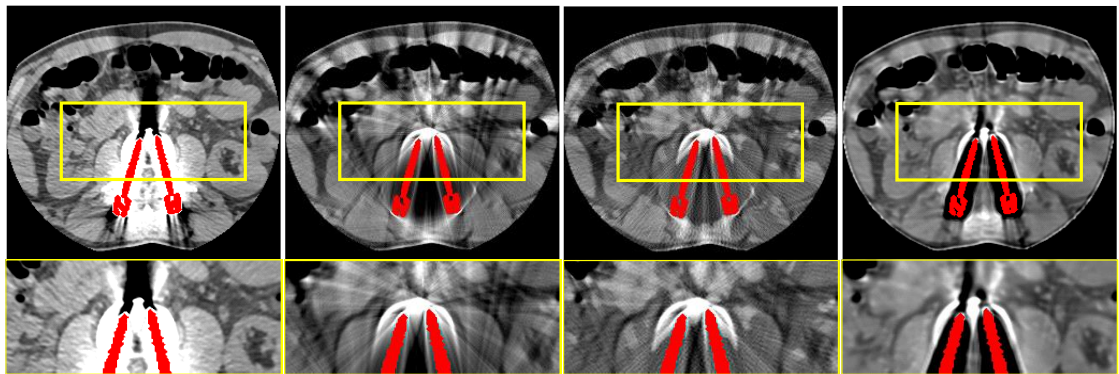
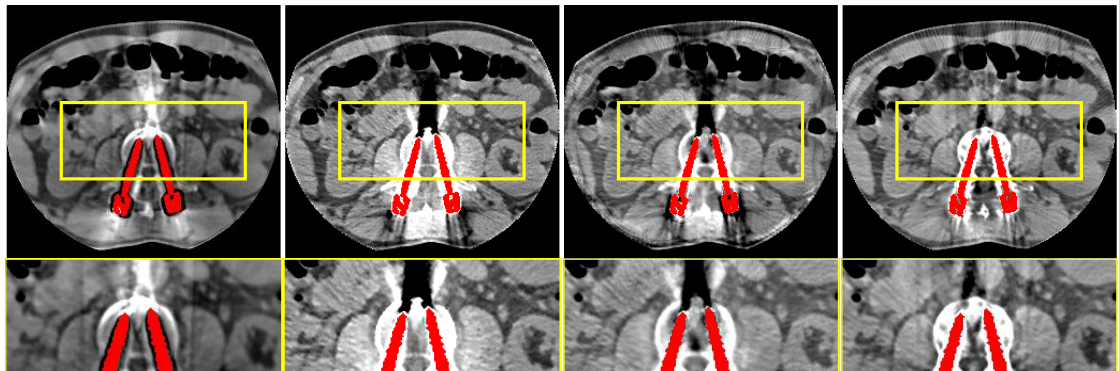

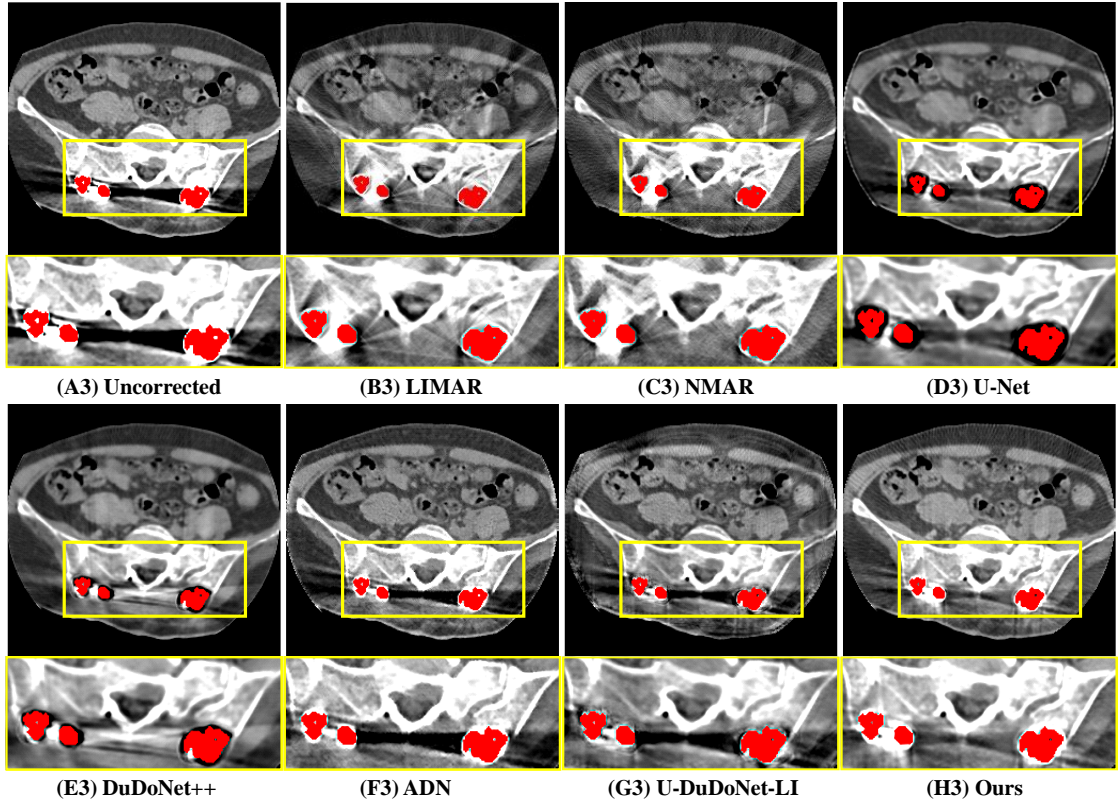

**Fig. 6.** Visual comparisons of different methods on clinical CT images of lumbar with small metal implants (A1), lumbar with large metal implants (A2), and pelvis (A3). The MAR images of LIMAR (B1-B3), NMAR (C1-C3), U-Net (D1-D3), DuDoNet++ (E1-E3), AND (F1-F3), U-DuDoNet-LI (G1-G3), and our method (H1-H3) are shown. The display window is [-110 560] HU and metal implants are colored in red for better visualization.

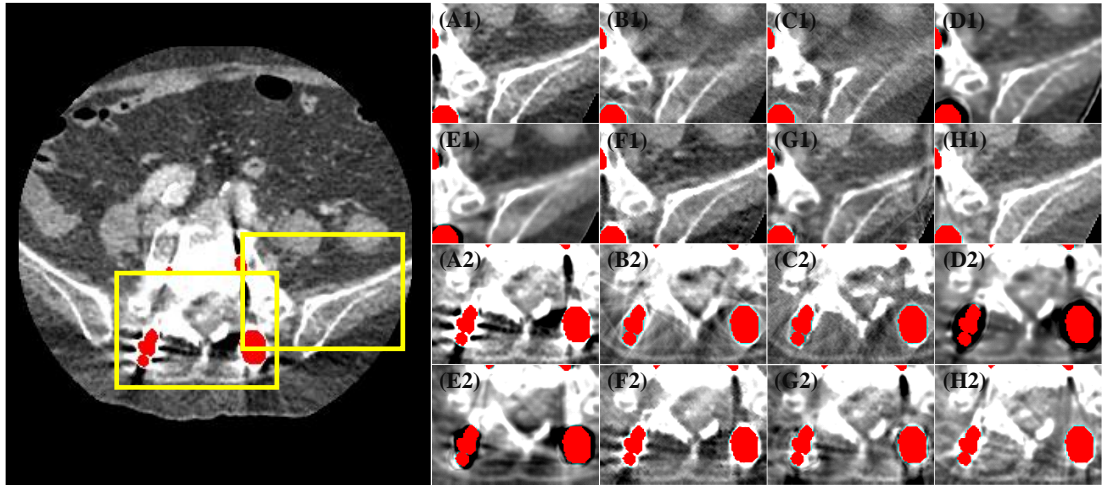

**Fig. 7.** A clinical case with real metal artifacts and the details of different SOTA methods results. The zoomed figures show uncorrected image (A1&A2), LIMAR (B1&B2), NMAR (C1&C2), U-Net (D1&D2), DuDoNet++ (E1&E2), AND (F1&F2), U-DuDoNet-LI (G1&G2) and our PND-Net (H1&H2). The display window is [-110 560] HU and metal implants are highlighted in red.

## C. Ablation Study

To evaluate the effectiveness of different components of our model, we design different

ablation models with the following modules:

**M1) NSD-Net:** The sinogram completion module with NSD-Net.

**M2) IR-Net+sinogram enhancement network:** Dual-domain learning with IR-Net and the sinogram enhancement network proposed in U-DuDoNet.

**M3) NSD-Net+IR-Net:** Dual-domain learning with NSD-Net and IR-Net.

**M4) NSD-Net+IR-Net+$\mathcal{L}_{IC}$:** Dual-domain network with the internal consistency loss.

**M5) NSD-Net+IR-Net+F-Net:** Dual-domain network with F-Net in artifact synthesis path.

**M6) NSD-Net+IR-Net+F-Net+$\mathcal{L}_{IC}$:** Our full model.

As summarized in Table 2, It is observed that M3 achieves higher SSIM and PSNR, and lower RMSE than M1, suggesting that IR-Net can facilitate image restoration. Fig. 8(D&E) shows that our proposed NSD-Net reduces more artifacts compared with the inpainting method proposed in U-DuDoNet, validating the effectiveness of non-local sinogram decomposition. Compared to M3, M4 attains better performance due to the capability of the IC loss, which can also be proved by comparing M5 and M6. The quantitative comparison of M3 and M5 indicates that F-Net can improve the MAR result, and our full model achieves the best performance among different ablation models.

Table 2. Quantitative evaluation of different ablation models on simulation data.

|  | M1 | M2 | M3 | M4 | M5 | M6 |
|---|---|---|---|---|---|---|
| SSIM | 0.734 | 0.937 | 0.945 | 0.947 | 0.948 | 0.949 |
| PSNR (dB) | 28.39 | 31.74 | 32.81 | 33.51 | 33.45 | 34.01 |
| RMSE (HU) | 46.77 | 31.51 | 27.53 | 25.02 | 25.72 | 24.12 |

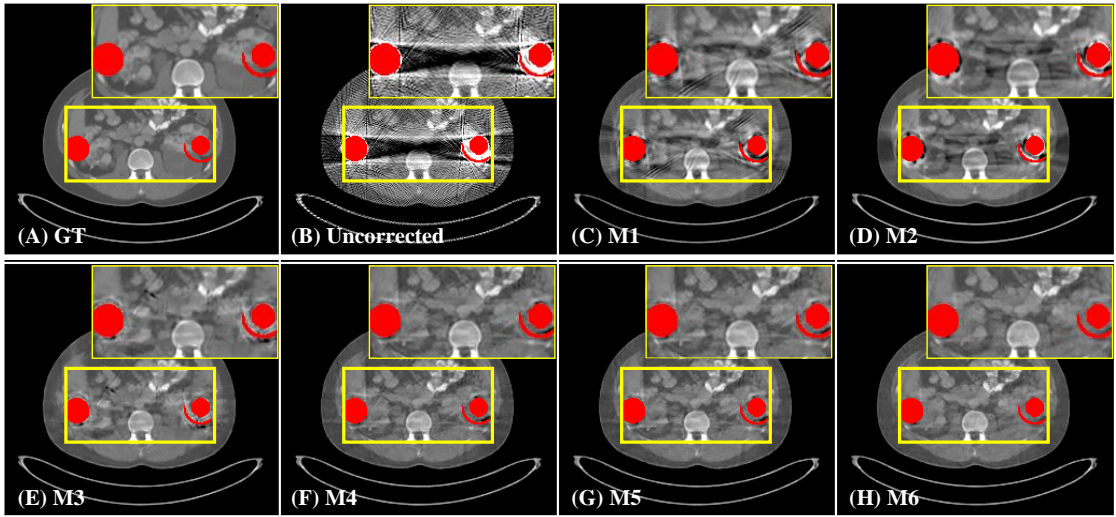

**Fig. 8.** Visual comparison of different ablation models of our method. The display window is [-320 600] HU and red marks stand for metal implants.

## V. DISCUSSION AND CONCLUSION

As many deep learning methods for metal artifact reduction have been proposed in recent years, this persistent and challenging problem in CT imaging is expected to be better addressed in the future. Since metal artifacts are structured in the image domain, many sinogram completion based methods are developed to restore the missing projection values in the metal trace region. However, the inconsistency between original projections and estimated values at the boundary of the metal trace can introduce secondary artifacts. Therefore, prior sinogram [21] and image post-processing [24, 28] methods are successively presented to achieve better MAR performance. Supervised dual-domain approaches, such as DuDoNet++, perform well on simulation data but attain poor performance in the real scenario due to the lack of paired clinical data. To synthesize paired training data, a common strategy is to artificially insert metal implants into clinical clean CT images to acquire metal-corrupted CT data [24]. In real clinical scenarios, the HU value of metal implants tends to saturate due to the limited CT range of the CT scanner, and such HU saturation leads to the low discrimination between metal implants and normal tissues. As a result, metal artifact reduction on real data is much more challenging than synthesized metal-corrupted CT images. Inspired by CycleGAN, an artifact synthesis path is generally incorporated for unpaired learning to improve the network performance on real clinical CT images. Following this idea, we propose an unsupervised dual-domain method to enhance the generalization of our model. Unlike U-DuDoNet [30] that directly adds artifact components to an arbitrary clean CT image, we design a fusion network to synthesize the metal-corrupted image since the artifact component in the image domain contains structural textures. U-DuDoNet solely trains a self-learned prior net with artificial metal-corrupted sinograms to guide the sinogram enhancement, whereas our method employs the simple but robust method LIMAR to produce a prior sinogram and Gaussian blur operation is utilized to smooth the LI sinogram. More importantly, we mathematically prove that beam hardening artifacts are mainly located within metal trace region, and employ MC simulation to verify that other artifact components are non-local in the sinogram domain and inevitable for MAR task. Different from previous methods [22, 23, 26, 30, 35] that only restore the corrupted sinogram within metal trace, we propose a novel non-local sinogram decomposition network based on this physical property of metal artifacts to acquire the weighted artifact component, and ablation experiments have demonstrated the effectiveness of such non-local sinogram decomposition strategy. In our method, we design an image restoration network IR-Net to reduce the residual and secondary artifacts without the loss of normal tissues. That is, the output of IR-Net is expected to be identical with the original input when IR-Net takes clean CT images as input. Therefore, we further propose an internal consistency loss to ensure the data fidelity of anatomical structures in the image domain. Quantitative comparisons in Table 1 show the effectiveness of such loss function.

In this paper, we present a physics based non-local dual-domain network for metal artifact reduction in CT imaging, where NSD-Net, IR-Net, and F-Net are jointly trained to achieve non-local sinogram decomposition, image restoration, and artifact synthesis processes. Our method can effectively remove dark bands and bright borders due to the non-local sinogram decomposition strategy. Residual and secondary artifacts are suppressed in the image domain with the guidance of internal consistency loss, and the generalization and robustness of our method on clinical CT images are improved via unpaired learning with a fusion network. Experimental results demonstrate that our method can achieve comparable MAR performance with state-of-the-art supervised methods on simulation data, and has potential to be utilized in clinical workflow.

**ACKNOWLEDGEMENTS**

This work was supported by National Key R&D Program of China (2022YFA1204203), National Natural Science Foundation of China (61971463, 82272131, 82202960), Ministry of Science and Technology Planning Project of Guangdong (208093178039).